# Quasiparticle dynamics in ballistic weak links under weak voltage bias: An elementary treatment

HERBERT KROEMER[*]

*Department of Electrical and Computer Engineering,
and QUEST, Center for Quantized Electronic Structures,
University of California, Santa Barbara CA 93106, USA*

A simple one-dimensional model for SNS weak links in the ballistic limit is presented. In the presence of a bias voltage, the quasiparticle state at any given instant of time is described as a superposition of that particular set of phase-dependent Andreev bound states that belongs to the specific phase difference present at this instant between the superconducting banks. The treatment—basically a form of adiabatic perturbation theory—has a strong formal similarity to the treatment of the $k$-space dynamics of an electron in a periodic potential under perturbation by an external electric field, sufficiently strong to cause transitions across the energy gaps between bands (Zener tunneling). It is shown that the quasiparticle wave function retains its phase information during analogous transitions between Andreev bands. The experimental observation of Shapiro steps at one-half the canonical voltage follows naturally from the model, along with some of the experimental properties of these steps, especially their much weaker temperature dependence, compared to the canonical steps.

**Key words:** Andreev reflections, ballistic, Shapiro steps, weak links.

## 1. Introduction: The problem

In a 1985 paper [1], Kümmel and Senftinger (KS) studied the time evolution of a quasiparticle (QP) wave packet in an idealized SNS weak link, under the influence of a weak external voltage bias, making the idealizing assumptions of perfectly ballistic transport, perfectly transparent SN interfaces, and zero Fermi velocity mismatch. The authors showed that, as a result of multiple Andreev reflections (ARs), the QP would pick up kinetic energy from the applied bias, until it was ejected from the Andreev well, into the downstream superconducting bank. However, the paper did not address questions of the ac Josephson effect under voltage bias, in the presence of this energy pickup.

---
[*] EMail: kroemer@ece.ucsb.edu

Experimentally, a pronounced—and highly anomalous—ac Josephson effect in ballistic SNS weak links has recently been reported by Drexler et al. [2] and Lehnert et al. [3, 4] (DL). In the present paper, we re-examine the KS treatment and attempt to reconcile it—at least qualitatively—with these experimental observations, and especially with the very pronounced anomalies found by DL.

In DL, the authors investigated the Shapiro steps induced, by high-frequency irradiation, in the dc current-voltage characteristics (CVCs) of superconducting weak links, which were based on InAs quantum wells as a coupling medium between Nb electrodes. The two investigations differ in the details of the device structure as well as the measurement technique, but both studies revealed a common behavior quite different from that in more conventional Josephson devices:

(a) In addition to the "canonical" Shapiro steps at the voltage $V = \hbar\omega/2e$, where $\omega$ is the irradiation frequency, the devices also showed strong steps at one-half that voltage,

$$V_{1/2} = \hbar\omega/4e, \tag{1}$$

indicating the presence of a strong component in the ac Josephson current at the frequency $4eV/\hbar$, twice the canonical Josephson frequency $\omega_J = 2eV/\hbar$.

(b) With increasing temperature, both kinds of steps decreased, but the half-integer steps did so much more slowly, persisting to higher temperatures than the integer steps, into a temperature range close to the critical temperature $T_c$ of the superconducting Nb banks, where indications of both the dc Josephson effect and the integer step had all but disappeared.

(c) By varying the drive frequency over a wide range, Lehnert found that the half-integer steps became more pronounced with increasing frequency.

Perhaps the most surprising of these observations is the temperature dependence; as pointed out by Lehnert et al., it rules out many potential explanations one might otherwise offer.

It was shown by Argaman [5, 6] that the observations can be explained in terms of a certain non-equilibrium model: In systems with long energy relaxation times, the voltages necessary to reach the Shapiro steps are sufficiently large to drive the quasiparticle (QP) energy distribution out of equilibrium, leading to a distribution function that contains itself a component oscillating with the canonical Josephson frequency. This ultimately causes the current to contain a component oscillating with twice the Josephson frequency. The higher the drive frequency, the larger the Shapiro step voltage, and hence the larger the non-equilibrium component, thus immediately explaining observation (c) above. Beyond that, Argaman's theory makes several quantitative predictions, essentially all of which were confirmed experimentally by Lehnert [3, 4]. For details of the theory, we must refer to the original papers [5, 6], which also give extensive references to related theoretical work by others. Of those other theoretical papers the ones most relevant to the present work are those by Averin and Bardas [7, 8], who consider related problems for superconducting quantum point contacts, drawing on a significantly different formalism.

The work by Argaman draws on the well-developed theoretical formalism for *diffusive* weak links, into which he incorporates non-equilibrium effects via a dis-equilibrated distribution function of the QPs over a quasi-continuum of Andreev bound states. However, the devices investigated were actually closer to the ballistic limit. Argaman argues—correctly—that the underlying physics should carry over to ballistic devices. In fact, the non-equilibrium effects should be more pronounced in the ballistic limit, where the perturbation by random processes is much weaker. It might therefore be useful to approach the same non-equilibrium physics from the opposite end, the purely-ballistic limit. Here, the only scattering processes considered are normal scattering and Andreev scattering at the super-normal interfaces, with scattering processes *inside* the normal material being neglected, or at best treated as a weak perturbation. This is of course again an over-simplification, albeit one in the opposite direction from the diffusive limit.

When a small external bias is applied to a ballistic weak link, two processes take place, one time-periodic, the other time-monotonic (non-periodic):

(a) The time-periodic process is the conventional ac Josephson current, just as in Josephson tunnel diodes. Its most obvious "fingerprint" is the occurrence of Shapiro steps in the dc CVC under irradiation with a high-frequency signal (see, for example, Tinkham [9])

(b) In ballistic structures in which multiple Andreev reflections (ARs) can occur, the *additional* process studied by can occur, where the quasiparticles may pick up energy from the bias field, in a way that is not oscillatory in time. The most obvious fingerprint of this phenomenon is the sub-harmonic gap structure often observed in the current-voltage characteristics (CVCs) of ballistic weak links (see, for example, Klapwijk et al, [10]).

Each phenomenon by itself has been discussed extensively in the literature, at various levels of sophistication. Our objective here is to give a *simple* unified treatment that treats both phenomena on a common basis, but on a more elementary level than what appears to be available in the literature.

Our treatment differs from that of KS in two ways: (a) We drop the restriction to perfectly transparent SN interfaces and zero Fermi velocity mismatch. (b) Rather than explicitly following the time evolution of a *localized* QP wave packet, we treat the problem in the spirit of adiabatic perturbation theory, in which the time evolution of an *extended* state is viewed as that of a linear superposition of states from a *time-dependent* set of Andreev bound states. Our treatment implicitly assumes a Bogoliubov-de-Gennes (BdG) hamiltonian as in KS, in which the dc voltage bias has been included, not through a conventional (time-independent) scalar potential, but through a time-dependent vector potential. The resulting BdG hamiltonian depends on time parametrically, making the problem readily tractable as an adiabatic perturbation problem. However, we shall not find it necessary to invoke the BdG equations explicitly.

Instead, we draw on a very close formal similarity to the dynamics of an ordinary electron in a periodic potential, under the influence of an applied electric field that is sufficiently strong to cause inter-band transitions (in semiconductor physics commonly referred to as Zener tunneling). Such a treatment leads to a very simple theoretical description of the physics of the basic phenomena, including the anomalies listed earlier.

The experimental examples coming closest to our ballistic limit are probably those SNS weak links in which the normal conductor is the high-mobility 2-dimensional electron gas in a semiconductor quantum well, like the InAs-based quantum wells studied at UCSB and elsewhere (for complete references, see Thomas et al. [11]). But our treatment itself should not be viewed as being specifically directed towards those devices.

## 2. Andreev bands

Our point of departure is the fact that, in the absence of an external bias voltage, it is possible to define a discrete set of current-carrying Andreev bound states, as first discussed by Kulik [12], and to express the flow of any current in terms of the occupancy of these states. The states come in pairs, with opposite currents for the two states of a pair. In the absence of a phase difference between the two superconducting banks, the two states of each pair are degenerate. In thermal equilibrium they then have the same occupation probability, and their currents cancel. If a phase difference $\varphi$ is present (but no *voltage* bias), this degeneracy is lifted. The states then no longer have equal occupation probabilities, and a net current can flow, even in the absence of a voltage bias. This is the equilibrium Josephson current.

When a *voltage* bias is applied, the Andreev bound states are no longer stationary eigenstates of the problem; they nevertheless remain central to the theory, as complete sets of basis states in terms of which to express the time-dependent actual quantum states as superpositions of Andreev bound states. This is the procedure we follow here.

## 2.1. Zero-barrier limit

The Andreev bound states have been extensively studied in the literature; the simplest and most widely studied case is that of a purely ballistic weak link with perfectly transparent S-N interfaces, containing neither a barrier nor a mismatch in Fermi velocity; the only difference between the superconductor and the normal conductor is that the former has a non-zero pair potential $\Delta$, while in the latter the pair potential is assumed to be zero. Because of our emphasis on simplicity, we restrict ourselves further to a one-dimensional problem. The energies of the Andreev bound states in this idealized limit were first given by Kulik [12]. These energies depend on the phase difference $\varphi$ between the two superconducting electrodes; Kulik's result may be expressed via the simple relation

$$E = E_c \cdot \left[ m - \frac{1}{\pi} \arcsin\left(\frac{E}{\Delta}\right) \pm \frac{\varphi}{2\pi} \right]. \tag{2}$$

Here, $E$ is the energy relative to the chemical potential, $m$ is a positive half-integer,

$$m = \tfrac{1}{2}, \tfrac{3}{2}, \ldots \text{ etc.}, \tag{3a}$$

$\Delta$ is the pair potential of the superconducting banks, and

$$E_c = \frac{\pi \hbar v_F}{L} \tag{4}$$

is an energy characterizing the coupling strength of the weak link in terms of the Fermi velocity $v_F$ of the semiconductors and the separation $L$ between the superconducting banks.

In addition to the positive-energy states *above* the chemical potential, there also exists a second set of states at the exact mirror energies *below* the chemical potential. We may include both sets in (2) by simply extending the $m$-values in (2) to include negative half-integers:

$$m = \pm\tfrac{1}{2}, \pm\tfrac{3}{2}, \ldots \text{ etc.} \tag{3b}$$

If one plots the energy levels described by (2) as functions of the phase difference, the states evidently fall into two sets with opposite slopes. Because the current is related to the slope via

$$I = \frac{2e}{\hbar} \cdot \frac{dE}{d\varphi}, \tag{5}$$

the two sets represent opposite currents; the upper sign in (2) belongs to states with positive currents, the lower sign to negative currents.

### 2.2. Andreev bands and gaps

The two sets of states described by (2) intersect: States belonging to opposite currents and to $m$-values that differ by $\Delta m$, will cross each other when $\varphi = \pi \cdot \Delta m$. This crossing-over of the states is unrealistic: it is a consequence of the extreme oversimplification of the model employed so far, which contains no mechanism by which forward-traveling states can be scattered into backward-traveling states, and vice versa. Almost any perturbation will cause the cross-over degeneracies to be removed, and lead to *Andreev bands*, separated by energy gaps (Fig. 1).

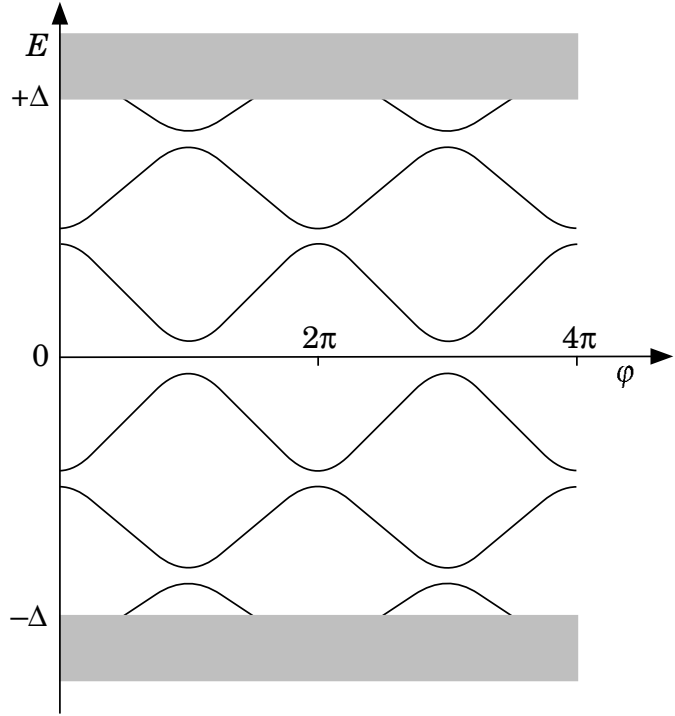

**Fig. 1.** Schematic Andreev bands, separated by gaps, as functions of the relative phase $\varphi$, in a weak link structure with non-negligible scattering between positive- and negative-current states. The graph is for a weak link of intermediate length $L$ such that there are two full bands each both above and below the chemical potential, with a third band merged partially into the continuum above and below the Andreev well. Note that we are deliberately *not* restricting the phase difference to an interval $2\pi$ wide, but employ an extended phase space.

In the literature, such *E* - *φ* diagrams are almost invariably discussed with the phase difference *φ* being restricted to an interval 2π wide. For our purposes—we will later wish to replace the phase *φ* with the time *t*—it will be more useful to discuss the *E* - *φ* diagram in *extended phase space*, where *φ* may assume arbitrary values. This is similar to the way the wave number *k* may assume arbitrary values in extended-*k*-space discussions of the dynamics of an electron in a periodic potential, a similarity to which we will return shortly.

Reflection barriers are often modeled by delta function potentials added to the interface, with a certain dimensionless strength parameter *Z* (see, for example, Blonder et al. [13]) If those potentials are sufficiently weak, the resulting gaps are easily obtained by perturbation theory, which yields gaps proportional to *Z*. We will not follow this line of thought here; for our purposes it is more useful to work with the gaps directly—whatever their origin—rather than in terms of a specific perturbation causing the gaps.

## 3. Interband transitions

In the presence of a bias voltage *V*, the phase difference *φ* between the two superconducting banks evolves according to the fundamental Josephson relation

$$\frac{d\varphi}{dt} = \frac{2eV}{\hbar}. \qquad (6)$$

Associated with each value of the phase difference is a separate set of Andreev bound states, and the central idea of our treatment is to use a time-dependent set of basis states in such a way that, at every instant of time, the overall state may be expressed as a superposition of those Andreev bound states that corresponds to the specific phase difference at that instant of time.

Consider now one Andreev bound state $|N, \varphi\rangle$ belonging to the specific Andreev band *N*, and within that band to the specific phase difference *φ*. In the absence of a bias voltage, *φ* itself would be time-independent, and the state $|N, \varphi\rangle$ would be a true stationary state of a quasiparticle. Assume next that, at time *t* = 0, a non-zero bias voltage *V* is turned on. According to (6), the phase difference *φ* will then change with time, and the initial state considered will evolve into a different state.

If the applied bias is sufficiently weak, and the gaps between the Andreev bands are sufficiently wide, the evolution of the state may then be described, at least to the first order, as an *adiabatic* change, by simply letting the phase difference *φ* evolve according to (6), while staying within the same band, without crossing the gaps. The energy of this time-dependent state evidently oscillates, with the "canonical" Josephson frequency

$$\omega_J = \frac{d\varphi}{dt} = \frac{2eV}{\hbar}. \tag{7}$$

Because of (5), the current also oscillates with the same frequency. This is of course simply the "ordinary" ac Josephson effect.

Once $\varphi$ becomes time-dependent, the states $|N, \varphi(t)\rangle$ are no longer exact solutions of the time-dependent BdG equations. In addition to the adiabatic motion of the state through $\varphi$-space within the initial Andreev band, there will also be transitions across the gaps to other bands, and it is these transitions that are responsible for the energy pickup of a quasiparticle in the presence of multiple ARs, up or down what we would like to call the *Andreev ladder*.

Formally, the problem is almost exactly the same as the problem of the $k$-space dynamics of an electron in a periodic potential, under the influence of an external applied uniform and time-independent force $F$. In this case, it is well known that, for a sufficiently weak force and sufficiently wide gaps separating the bands, the electron dynamics can be described by the familiar relation

$$\hbar \frac{dk}{dt} = F, \tag{8}$$

which is evidently analogous to (6), with the substitutions $k \leftrightarrow \varphi/L$ and $F \leftrightarrow 2eV/L$.

However, at the same time, transitions to other bands also take place. The instantaneous transition probability increases roughly exponentially with decreasing energy separation between the two bands involved, taken at the particular $k$-value that is present at that instant. The probability reaches a sharply peaked maximum at that point in $k$-space at which the energy separation goes through a minimum, the net energy gap. The theory, found in advanced semiconductor texts [14], shows that the integrated transition probability for a single passage through the gap region is given by

$$T_k \approx \exp\left(-\frac{\pi}{4}\sqrt{\frac{2m^*}{\hbar^2}} \cdot \frac{E_G^{3/2}}{F}\right), \tag{9}$$

where $m^*$ is an the effective mass in the two bands, taken at the narrowest gap (and assumed to be equal for both bands). This is what in semiconductor physics is referred to as *Zener tunneling*; the analogy of the QP dynamics to Zener tunneling has been noticed previously by Averin and Bardas [7, 8] for the somewhat different case of superconducting quantum point contacts.

The $k$-space result (9) may be transferred directly to our $\varphi$-space problem by the substitutions

$$\frac{\hbar^2}{m^*} = \frac{d^2E}{dk^2} \to L^2 \frac{d^2E}{d\varphi^2} \approx \frac{1}{2}L^2 \cdot E_c \tag{10}$$

and

$$F \to 2eV/L. \tag{11}$$

The result may be written as

$$T_\varphi \approx \exp\left(-\frac{\pi}{4} \cdot \sqrt{\frac{E_G}{E_c} \cdot \frac{E_G}{eV}}\right). \tag{12}$$

For a sufficiently narrow gap, say, $E_G = E_c/4$, and a bias voltage of, say, $V = E_G/e$, we find a transition probability $TKO = \exp(-\pi/8) = 0.68$, large, but distinctly less than 100%.

Inasmuch as the $k$-space result (9) is itself only an approximation, valid under assumptions that might not carry over to the $\varphi$-space case, (12) is also only an approximation. But it indicates the correct trends one should expect with respect to the dependence of the transition probability on key parameters such as the gap between Andreev bands, and the bias voltage.

Taken as an indication of trends, (12) shows that the transfer to higher energies decreases rapidly with increasing gap, but for a given finite gap, it increases rapidly with increasing bias voltage. The case discussed by KS is the specific case of zero gap; then the quasiparticle crosses over to the adjacent band with 100% probability, but this is clearly an unrealistic limit.

## 4. Interband ac Josephson effect

### 4.1. Current dynamics

In the case of $k$-space dynamics, one of the key properties of the interband transfer is that an electron in a Bloch wave with a sharp (traveling) value of $k$ will make transitions to another band only to states with the same (traveling) $k$ in that band [15, 16]. This property carries over to our problem (Fig. 2).

Evidently, the state into which the transfer takes place, carries exactly the same phase information as the original state. As a result, there is no loss of phase information, and hence such transfer will not destroy the ac Josephson effect!

However, the transitions drastically alter the *details* of the ac currents: For any given value of $\varphi$, adjacent bands carry currents in with opposite sign. Hence, an interband transition will cause a reversal of the current direction relative to what would have happened if there had been no transition.

Conversely, the current direction will remain the same as what it was before the transition, while in the absence of the transition it would have reversed.

This behavior naturally leads to a strong ac component oscillating at $2\omega_J$, twice the Josephson frequency $\omega_J$. As a simple example of how this comes about, consider specifically a QP in the highest band just below the chemical potential ($m = -1/2$), and assume that the phase difference is initially $\varphi = 0$ (Fig. 3). In the absence of interband transition, this state would remain in its initial band as the phase evolves linearly in time, and the QP's energy, as well as its contribution to the overall current, would oscillate with the Josephson frequency $\omega_J$. There would be no dc current.

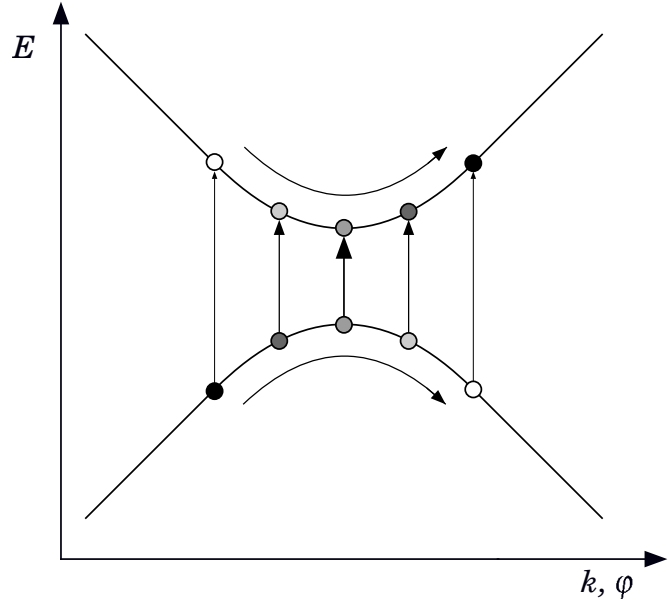

**Fig. 2.** Interband transitions always involve pairs of states with the same traveling $k$- or $\varphi$-value.

But if, during crossing the region of narrowest gap, around $t = t_\pi = \pi/\omega_J$, the QP makes a transition to the next higher band ($m = +1/2$), the current will not reverse following the transition. Instead, it will only dip to zero, and then recover to a positive value, at least until $t = 2t_\pi$ (Fig. 3). The net result is that this time interval from $t = t_\pi$ to $t = 2t_\pi$ will make both a dc contribution to the overall current, and a strong ac contribution at $2\omega_J$, twice the Josephson frequency!

What happens subsequently depends on the details of the Andreev band structure, which in turn depends strongly on the length $L$ of the weak link (see sec. VI below).

In weak links that are sufficiently short that there is no higher Andreev band inside the Andreev well ($E < \Delta$) at $\varphi = \pi$, there will probably be a high probability of ejection of the QP from the Andreev well into the downstream superconducting bank, and only a low probability for the QP to stay in the band and return to lower energy. This is the case especially favorable for a strong $2\omega_J$ current component. The subsequent return to lower energies would largely cancel the $2\omega_J$ and only leave a $\omega_J$ component in that fraction of events where the QP would not be ejected.

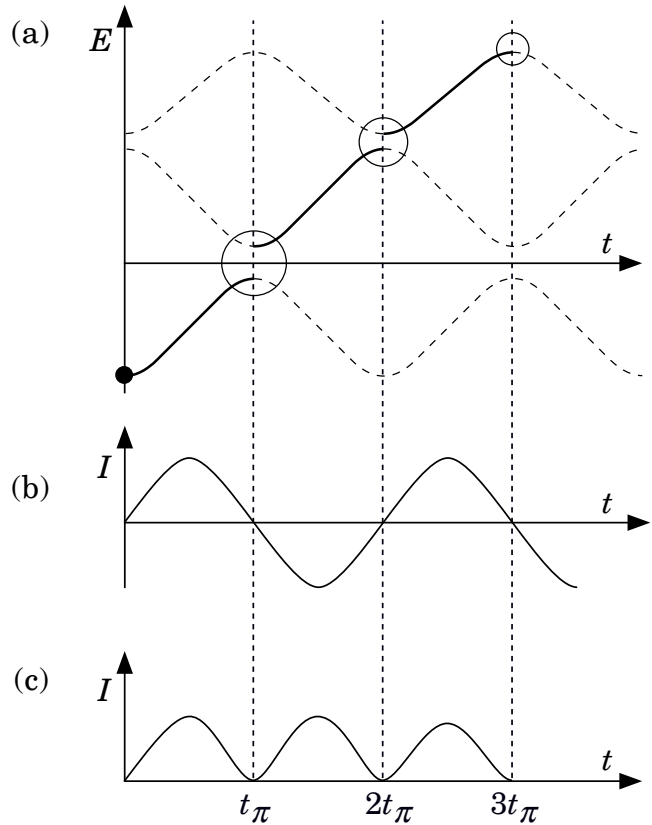

**Fig. 3.** (a) "Andreev ladder" for a QP starting at the point marked •, and making a succession of interband transitions (circles). (b) Current contribution vs. time if the QP stayed within its initial Andreev band. (c) Current contribution vs. time in the presence of the interband transitions shown in (a). The rounding of the current minima occurs because the transition is not abrupt at phase multiples of $\pi$, but takes place over a finite phase range (see Fig. 2). The current amplitude decreases as the bands get narrower. The time $t_\pi$ is defined as $t_\pi = \pi/\omega_J$, one-half the canonical Josephson period.

The opposite limit of a long weak link with a large number of Andreev bands, is somewhat different. If the interband transition probability is high, the QP, upon reaching $\varphi = 2\pi$, is likely to make another transition to the next higher band ($m = 3/2$). The contribution of a dc current plus an ac current at $2\omega_J$ would then continue. For a sufficiently long weak link, with many Andreev bands, one might visualize a (low-probability) extreme limit of an unbroken sequence of upward transitions (an *Andreev ladder*) whenever $\varphi$ crosses a multiple of $\pi$, until the QP is finally ejected from the Andreev well into one of the superconducting banks. Under these extreme idealized assumptions, there might in fact be no significant current contribution at $\omega_J$ itself during the process of "climbing up the Andreev ladder"

However, the appearance of a strong $2\omega_J$ component in this limit would be counteracted by the following. A near-100% transition probability would require the near-absence of gaps between the Andreev bands. This, in turn, would imply a reduction of the $E(\varphi)$ relation to the simple near-linear form (2). But for a linear relation, the current would be a pure dc current, without any ac current at *any* frequency, at least not from the lower Andreev bands. The principal exception would occur for the highest Andreev band, where the ejection from the Andreev well occurs periodically in $\varphi$, implying a weak ac Josephson current contribution. Such a weak ac Josephson current, presumably due to this remaining periodicity, is readily visible in the calculations by Jacobs et al. [17].

On balance, the occurrence of a $2\omega_J$ current component therefore requires a transition probability that is neither too large nor too small. But this also complicates a quantitative determination of the strength of the $2\omega_J$ current component, requiring an analysis of the current frequency spectrum resulting from a probabilistic mixture of (*i*) upward transitions, (*ii*) no transitions, and (*iii*) even downward transitions. As a result, the frequency spectrum of the Josephson current will be more complex; it will still contain the fundamental Josephson frequency $\omega_J$ associated with $\pi$-crossings without inter-band transitions. But there will always be a strong component at $2\omega_J$ associated with those crossings where transitions take place. Furthermore, inasmuch as the current is no longer strictly periodic in time, its frequency spectrum will be continuous rather than discrete, with significant broadening of the peaks at $\omega_J$ and $2\omega_J$, especially the latter. Such a broadening is clearly visible in the experimental data [3, 4]; an analysis of this broadening might provide important clues concerning the details of the QP dynamics.

I have not attempted a quantitative treatment of the consequences of the probabilistic mixture of transitions; I will return to this question below, in the *Discussion* section of this paper.

Before closing the topic of the $2\omega_J$ current, the following remark is in order. An essential ingredient in the occurrence of a net $2\omega_J$-current is the QP ejection into the superconducting banks once the QP energy reaches the pair potential $\Delta$ at the top of the Andreev well. If no such ejection took place (nor any inelastic collisions), any QP reaching one of the higher rungs of the Andreev ladder would ultimately have to climb back down on the ladder. The interband Josephson current associated with this downward climb is opposite to that with the upward climb. If both upward and downward climb were equally frequent, there would be no *net* interband current at *any* frequency, not even a dc contribution. In reality, some of the QPs climbing up the ladder will eventually reach the energy $\Delta$ at the top of the ladder, at which point they are ejected into one of the superconducting banks, with no compensating current due to downward climb afterwards. It is only because of the existence of QP ejection (and inelastic scattering, which has a similar effect) that there will be a dc current, and along with it an ac current at twice the Josephson frequency.

### 4.2. Shapiro Steps

As was pointed out already in the experimental papers [2-4], an ac Josephson current contribution at $2\omega_J$ will cause Shapiro steps to occur at one-half the canonical Josephson voltage, as is indeed observed. Furthermore, Lehnert et al. [3, 4] made the key observation that, by varying the drive frequency over a wide range, the half-integer steps become more pronounced with increasing frequency. Inasmuch as a higher drive frequency shifts the step voltage to higher values, such a dependence is exactly what our simple theory would predict: According to (12), an increase in bias voltage increases the transition probability to higher bands, and with a strong $2\omega_J$ current being a consequence of interband transitions, the observed result follows naturally. The same conclusion was already drawn by Argaman, by a slightly different, but ultimately related argument.

## 5. Temperature dependence

As stated in the *Introduction*, perhaps the most remarkable observations of DL was that of the drastic difference in the temperature dependence of the two kinds of Shapiro steps [2-4]. This observation, already explained by Argaman's non-equilibrium theory, also finds a very natural explanation in our treatment.

At $T = 0$, and in the absence of a voltage bias, the Andreev bands above the chemical potential are empty, while those below the chemical potential are fully occupied. The current contributions from the different full bands do not cancel, but lead to a net Josephson current of the familiar form $I = I_e \cdot \sin\varphi$, with the possibility of higher harmonics. The latter should be negligible in the limit of long weak links.

With increasing temperature, quasiparticles are transferred from negative-energy states to positive-energy "mirror" states, which carry a current opposite to that of the negative-energy states, partially canceling the current contributed by the latter. The net result is a decrease in the critical current $I_c$ with increasing temperature, roughly exponential in the temperature [12]. The longer the weak link, the more rapid the decrease.

When a voltage bias is applied, we must distinguish the current contributions from QPs that remain in their band and those that undergo interband transitions and climb up the Andreev ladder. In the absence of interband transitions, the net thermal occupation of each Andreev band would remain largely unchanged by applying a small voltage bias, especially at temperatures where the thermal energy $kT$ is significantly above the width of the bands. The only difference is that now the Josephson current becomes an ac current, but its amplitude $I_c$ remains essentially equal to the dc critical current appropriate to the temperature that is present. Which of course means that the ac Josephson current at the fundamental Josephson frequency decreases rapidly with increasing temperature—as do the fundamental Shapiro steps.

The situation is quite different for those QPs that undergo interband transitions and climb up the Andreev ladder, to bands that would have only a small thermal occupation probability. The occupation of these bands is then determined much more by the interband transition probability than by the temperature. Inasmuch as it is interband transitions that cause a strong ac current contribution at twice the Josephson frequency, it follows naturally that the $2\omega_J$ current should fall off less rapidly with increasing temperature. The dominant effect of increasing temperature should be the decrease of the gap parameter $\Delta$ of the superconducting electrodes.

## 6. Length dependence

We have so far ignored the effect of the inter-electrode separation $L$ on the device behavior. This effect is substantial, with important consequences, especially for the number of Andreev bands within the Andreev well, their widths, and that of the gaps between bands.

For reference, consider first the barrier-free limit discussed earlier, with the simple $E$-$\varphi$ relation (2). At $\varphi = 0$ the total number of states with energies $0 < E < \Delta$ is easily found to be given by

$$N(0) = 2 \cdot \text{int}\left(\frac{\Delta}{E_c}\right) + 2, \tag{13a}$$

where int($x$) is the largest integer smaller than $x$. At $\varphi = \pm\pi$, one obtains

$$N(\pi) = 2 \cdot \text{int}\left(\frac{\Delta}{E_c} + \frac{1}{2}\right) + 1, \tag{13b}$$

Note that $N(0)$ is an even number, while $N(\pi)$ is odd, differing from $N(0)$ by ±1. However, these simple relations apply only to the case of vanishing Andreev gaps. In the presence of gaps, the top bands might get pushed out of the well, whereas the next-lower bands might get pushed deeper into the well.

For comparison with available experimental date, it is instructive to apply these considerations to the devices studied by Lehnert et al. [3, 4]. These devices had high electron sheet concentrations in the upper-$10^{12}$cm$^{-3}$ range. If one takes the strong non-parabolicity of InAs into account, such concentrations imply a Fermi velocity of about $1\times10^8$cm/s [18]. If one naively uses this value, along with Lehnert's electrode separation $L = 1.2\mu$m, one estimates $E_c \approx 1.7$meV, a value larger than the pair potential $\Delta = 1.5$meV appropriate for high-purity Nb at low temperatures. This would imply two positive-$E$ Andreev bands at $\varphi = 0$ and three at $\varphi = \pi$. The development of a significant gap between the Andreev bands would probably push the uppermost of these bands above $\Delta$, leaving just one band at $\varphi = 0$ and two at $\varphi = \pi$.

Evidently, even though the experimental devices studied by Lehnert et al. are long by comparison with the conventional superconducting coherence length, they are not long in Kulik's sense, of having a large number of Andreev bands confined to the Andreev well.

However, the above estimate is almost certainly too naive, and may overstate the case: The model presented here is one-dimensional, but real devices are at least two-dimensional. Even in a true 2-D quantum well, there is an angular distribution of velocities, and what matters would be the component of the Fermi velocity in the direction of current flow, which varies from its maximum value down to zero. Furthermore, because of the high electron sheet concentrations employed, at least one higher 1-D subband was occupied in these quantum wells. This further increases the uncertainty about which Fermi velocity to use. To allow for the two-dimensional nature of the transport, one might have to view the devices as a parallel combination of multiple channels with a continuous distribution of Fermi velocities from $1\times10^8$cm/s down. This would presumably call for an extension of our treatment along the lines of the work of Schüssler and Kümmel [19], greatly losing the simplicity of the present treatment in the process. I have not attempted such a treatment.

Regardless of these uncertainties, one as-yet-untested prediction of this model can be made: Just as the half-voltage Shapiro steps decrease less rapidly with increasing temperature than the normal steps, we would also expect them to decrease less rapidly with increasing length $L$: With increasing length, both the energy $E_c$ and the Andreev gaps widths $E_G$ should decrease. But for a fixed bias voltage, like the voltage associated with the Shapiro steps for a fixed drive frequency, the relation (12) then implies a larger transition probability, with its beneficial effect on the half-voltage

Shapiro steps. Such an effect may already have been present in Lehnert's devices, which had an unusually large electrode separation of about $1.2 \mu$m, much larger than what has been employed in most work with quantum-well-based weak links. This evidently calls for more detailed research on the length dependence of weak links, especially on devices with even larger electrode separations.

# 7. Discussion: Open issues

We have presented an elementary treatment of QP transport in long ballistic weak links under the influence of a weak voltage bias, but sufficiently strong to cause non-negligible QPs transfer out of each Andreev band into successively higher bands. The treatment gives a unified description of both the conventional equilibrium ac Josephson current with a fundamental oscillation frequency $\omega_J$, and of the non-equilibrium current with twice that frequency. In particular, it gives a natural explanation of the different temperature dependences of the two currents.

However, the model presented here is a "bare-bones" model, neglecting essentially all complications whose inclusion is not essential to the limited purpose of giving a *qualitative* explanation of the basic experimental facts. A more quantitative comparison with actual observation obviously calls for an elaboration of several of its aspects of the model, an elaboration that has not been undertaken yet.

We have discussed already the need for going beyond a one-dimensional model In the following, we address several scattering-related issues.

### 7.1. Basic scattering

We have also pointed out earlier the need for including the probabilistic competition between normal and Andreev reflections caused by the probabilistic nature of interband transitions. A simple rate equation treatment for this competition, in a somewhat different context, was given already before KS, in 1983, by Octavio et al. [20] . Unfortunately, that treatment does not address the question of the phase preservation so important for the understanding of the ac Josephson effect under these conditions.

We have completely neglected *inelastic* scattering in the present work. It should have consequences similar to the effects of QP ejection from the Andreev well; the question is: how important is it? Working with samples of somewhat lower electron concentration than the UCSB samples, Morpurgo has estimated the inelastic mean free path as about $15 \mu$m at 1.7 K [21]; similar values are obtained by performing a similar analysis on (unpublished) data by Thomas [18]. While these estimates suggest that inelastic scattering plays indeed only a role less than QP ejection in the comparatively short devices investigated by DL, it is by no means negligible weak, and its importance would increase in devices with larger electrode separation and/or smaller Fermi velocities. We consider the much longer inelastic mean free paths assumed in the theoretical work of Gunsenheimer and Zaikin [22] as unrealistically large, at least for the InAs coupling medium of interest here.

### 7.2. Diffusive Andreev scattering

There is an issue concerning the nature of the ARs themselves, especially under oblique incidence: The naive "textbook picture" of ARs at super-semi interfaces is one of specular retro-reflections in which the Andreev-reflected QP re-traces (almost) exactly the trajectory of the incident QP. However, the Groningen group has recently presented compelling evidence that this is not the case in the Nb-InAs-Nb weak links studied by them, and that the ARs are in fact diffusive, with the direction of the reflected-QP trajectory being almost random [23-25]. The origin of the diffusive nature of the scattering appears to be technological: In order to obtain super-semi interfaces of high transparency, oxide barriers must be eliminated, which is done by in-situ low-energy Argon sputter cleaning of the InAs surface just prior to the Nb deposition. While this procedure has the desired effect of creating a high-transparency interface, it was shown by Magnée et al. that it also creates a high concentration of defects inside the InAs [26], which in turn lead to diffusive ARs. There are many differences between the UCSB devices (including those of DL), and those of the Groningen group, especially in lithographic dimensions and, to a lesser extent, in the electron sheet concentrations in the 2-D electron gas. However, both groups use essentially the same MBE layer structure and interface cleaning procedure. Hence, the diffusive ARs are almost certainly also present in the UCSB devices, and there can be no doubt that they must be taken into account in any attempt to understand QP dynamics under bias.

### 7.3. Inter-subband scattering

Finally, there is a puzzling open question concerning the potential role of inter-subband scattering, at least in the UCSB devices. Essentially all *published* UCSB devices—but apparently none of those of other groups—employed electron sheet concentrations above $5 \times 10^{12} \text{cm}^{-2}$, up to about $8 \times 10^{12} \text{cm}^{-2}$. This choice was made deliberately, in response to the observation, made already by Nguyen [27], and confirmed by Thomas [18], that such high concentrations led to much stronger superconductivity effects than sheet concentrations in the low-$10^{12} \text{cm}^{-2}$-range, even in the presence of a significant reduction of mobilities at the high concentrations.

The highest low-temperature (10-12 K) mobility we have observed was $9.8 \times 10^5 \text{cm}^2/\text{V} \cdot \text{s}$, at a sheet concentrations of $1.5 \times 10^{12} \text{cm}^{-2}$ [27], and values approaching this maximum have been seen repeatedly, at similar sheet concentrations (see, for example, Blank et al. [28]). By comparison, the mobilities in the more heavily doped published weak-link samples were between a factor 1/3 and 1/10 lower.

Calculations show that in all samples a significant occupancy of multiple subbands was present, which in turn implies inter-subband scattering (ISBS). Presumably, much of the mobility reduction is caused by ISBS rather than impurity scattering. This hypothesis is strongly supported by measurements of Nguyen [27] and of Koester [29] on gated Hall effect samples. Both authors showed that the mobilities, measured as functions of the sheet concentration, display a maximum at a concentration of about $1.5 \times 10^{12}$ cm$^{-2}$, and drop rapidly to less than half that value for concentrations significantly exceeding $2 \times 10^{12}$ cm$^{-2}$. This is precisely what one would expect for ISBS.

Considering that ISBS between *uncorrelated* electrons would be a phase-breaking process, it should be detrimental to superconductivity and hence to weak-link performance. Yet our experience has not borne this out: whatever detrimental effects ISBS might have, they are clearly outweighed by the beneficial effects of the increased sheet concentration.

Ultimately, this observation raises the question of whether the independent-particle concept of well-defined subbands remains, in fact, applicable to structures in which the presence of superconducting electrodes might introduce a phase correlation between the electrons. Such a phase-correlated interaction might re-normalize the subband. Quite possibly, a hybridization might occur, qualitatively similar to what has been observed in coupled double quantum wells by Davies et al. [30]. Admittedly, this is, at present, pure speculation, but at the very least it suggests a closer investigation into the ISBS puzzle.

## 8. Acknowledgments